\documentclass[a4paper,12pt,aps,superscriptaddress]{article}
\usepackage{amsmath}

\usepackage{amsfonts}
\usepackage{amssymb}

\usepackage{float}
\usepackage{epsfig}

\newcommand{\be}{\begin{equation}}
\newcommand{\ee}{\end{equation}}
\newcommand{\bea}{\begin{eqnarray}}
\newcommand{\eea}{\end{eqnarray}}

\begin{document}
\title{Integer Partitions and Exclusion Statistics}
\date{\today }
\author{A. Comtet$^{1,2}$ , Satya N. Majumdar$^{1}$ and 
St\'ephane Ouvry$^{1}$\\
$^{1}$Laboratoire de Physique Th\'eorique et Mod\`eles
Statistiques,\\ 
Universit\'e de Paris-Sud, CNRS UMR 8626, 91405 Orsay 
Cedex, France \\
$^{2}$ Institut Henri Poincar\'e, 11 rue Pierre et Marie Curie, 75005 
Paris, France\\
}

\maketitle





\begin{abstract}
We provide a combinatorial description of exclusion statistics in terms of minimal 
difference $p$ partitions. We compute the probability distribution of the number 
of parts in a random minimal $p$ partition. It is shown that the bosonic point $ 
p=0$ is a repulsive fixed point for which the limiting distribution has a Gumbel 
form. For all positive $p$ the distribution is shown to be Gaussian. 

\noindent

\vskip 0.2cm


\vskip 0.2cm


\end{abstract}

\section{Introduction} 
Integer partition problem has a long history going back to Euler. The classical question asks: in how many ways $\rho(E)$ 
can one partition
an integer $E$ into nonzero integer parts $E=\sum_j h_j $ such that
$h_j\ge h_{j+1}$ for all $j=1,2,\ldots$? For example, $\rho(4)=5$:  $4=4=3+1=2+2=2+1+1=1+1+1+1$. 
Pictorially, one can represent
the part
$h_j$ as the height of the $j$-th column with nonincreasing heights such that the total height under the
columns is $E$. Hardy and Ramanujan proved \cite{HR} that for 
large 
$E$,
$\rho(E) \simeq \frac{1}{4}\frac{1}{3^{1/2}E} {\rm
e}^{a \sqrt{E}}$, with $a=\pi \sqrt{2/3}$. Similarly, one can
ask the number of ways of partitioning the integer $E$ into 
{\it distinct} integer summands, i.e., $E=\sum_j h_j$ such that $h_j>h_{j+1}$ with strictly decreasing heigth. For example, 
the integer $4$ can be partitioned
into distinct summands in only two ways,  $4=4=3+1$. In this 
restricted case
it is known that asymptotically for large $E$~\cite{Andrews},
$\rho(E)\simeq \frac{1}{4} \frac{1}{3^{1/4}E^{3/4}} {\rm e}^{b \sqrt{ E}}$
where $b=a/{\sqrt{2}}=\pi/\sqrt{3}$.

Another way of representing integer partitions makes clear
the connection with a gas 
of noninteracting quantum particles. Let $n_i$ be the number of columns
of height $h=i$ in a given partition, i.e. the number of times the summand
$i$ appears in a given partition. For example, in the partition $4=2+1+1$, 
one has $n_1=2$, $n_2=1$ and $n_j=0$ for all $j>2$. Then, $E= 
\sum_{i} n_i \epsilon_i$
where $\epsilon_i=i$ for $i=1,2,\ldots$ represent equidistant single particle energy levels
and $n_i=0,1,2,\ldots$ represents the occupation number of the $i$-th
level. In the unrestricted problem, the occupation number
$n_i=0,1,2\ldots$ (Bosons) whereas in the restricted problem $n_i=0,1$ (Fermions). 
Therefore, $E=\sum_i {n_i} {\epsilon_i} $ is the total energy of the system and 
\begin{equation}
\rho(E) = \sum_{n_i} \delta\left(E-\sum_{i=1}^{\infty} n_i 
\epsilon_i\right).
\label{re1}
\end{equation}

If in addition, one restricts the number of summands to be $N$, then the number
$\rho(E,N)$ of ways of partitioning $E$ into $N$ parts is simply the micro-canonical
partition function of a gas of quantum particles
with total energy $E$ and total number 
of particles $N$
\begin{equation}
\rho(E,N) =\sum_{n_i} \delta\left(E-\sum_{i=1}^{\infty} n_i
\epsilon_i\right)\,
\delta\left(N-\sum_{i=1}^{\infty} n_i\right).
\label{ren1}
\end{equation}
Evidently, $\rho(E)= \sum_{N=0}^{\infty} \rho(E,N)$. Even though the sum $\rho(E)$
has similar asymptotic behavior for large $E$ for Bosons and Fermions, i.e.,
$\ln (\rho(E))\sim \sqrt{E}$ (up to a constant prefactor), we will show
in this paper that $\rho(E,N)$, as a function of $N$ for a fixed $E$, has
rather different behavior for Bosons and Fermions.

Thus a gas of non-interacting Bosons or Fermions occupying
a single particle equidistant spectrum ($\epsilon_i=i$) both
have a combinatorial interpretation in terms of partitions
of an integer $E$ into $N$ parts.

\begin{itemize} 
\item Bose statistics corresponds to the case of unrestricted 
partitions $n_i=0,1,2\ldots$.

\item Fermi statistics corresponds to the case of restricted partitions with distinct 
summands $n_i =0,1$.

\end{itemize}
A natural question, that we address in this paper, is  how to provide
a combinatorial description of a quantum gas obeying exclusion statistics.
Exclusion statistics is a generalization of Bose and Fermi 
statistics~\cite{Haldane,LLLanyons,Wu,MS}. It
 has been found explicitly in quantum models of interacting particle systems,
notably in the two dimensional lowest-Landau-level (LLL) anyon model \cite{LLLanyons}
(i.e. the anyon model projected into the LLL  of a strong magnetic field) and the 
one dimensional Calogero model~\cite{Calogero,Moser,Sutherland,Isakov,Poly,Haldane2}. Note that the Calogero model can be obtained as a particular limit of the LLL-anyon model~\cite{Ouvry}, the latter being a particular exactly solvable projection of the anyon model: it follows that  exclusion statistics  is  deeply rooted in the more general concept of anyon statistics~\cite{Leinaas}. Unlike the Bose 
and Fermi 
statistics which 
describes
noninteracting particles, a combinatorial description of exclusion statistics
is a priori quite nontrivial since the underlying
physical models with exclusion statistics describe truly interacting $N$-body
systems. 

We show in this paper that a combinatorial interpretation of exclusion statistics 
involves a generalization
of the partition problem known as  the minimal
difference partition (MDP) problem.  
In MDP, one partitions a positive integer E
into $N$ nonzero parts, $E=\sum_{j=1}^N
h_j $ (with $h_j>0$ for all $j=1,2,\ldots, N$) such that each summand exceeds the next
by at least an integer
$p$, i.e. $(h_j-h_{j+1})\ge p$ for all $j=1,2,\ldots ,N-1$.
Therefore, $p=0$ corresponds to
unrestricted partitions (Bosons) and $ p=1$ to restricted partitions (Fermions) 
into distinct parts.
Even though the parameter $p$ in MDP is an integer, one can analytically
continue the results to noninteger values of $p$ and
we will show that for $0<p<1$, the MDP corresponds to a gas of quantum particles
obeying exclusion statistics.

Apart from establishing this equivalence between the MDP problem
and exclusion statistics, we also provide a detailed analysis
of the asymptotic behavior of $\rho_p(E,N)$, i.e., the number
of ways the integer $E$ can be partitioned into $N$ parts
in the MDP problem, for all $p\ge 0$. This analysis tells
us how the variable $N$ fluctuates from one partition to another
for fixed $E$. Indeed, defining $\rho_p(E)=\sum_N \rho_p(E,N)$
as the total number of partitions of $E$ and treating all
such partitions equally likely,  the ratio
$P_p(N|E)= \rho_p(E,N)/\rho_p(E)$ is the 
probability distribution of the random variable $N$, given $E$.
We show that this distribution, properly centered and scaled,
has rather different limiting shapes for $p=0$ and $p>0$.
While for $p=0$ the scaled distribution is asymmetric 
and has a Gumbel shape, for $p>0$ (including the Fermionic case $p=1$)
the scaled distribution is symmetric and has
a Gaussian shape.

At this point, it may be useful to summarize our main mathematical results
for the asymptotic behavior of $P_p(N|E)$. For the Bosonic case ($p=0$),
the limiting shape of the distribution was first derived by Erd\"os and Lehner
using rigorous methods involving upper and lower bounds~\cite{Erdos}. In this paper
we calculate the limiting shapes of $P_p(N|E)$ for all $p\ge 0$. Moreover,
our method allows us to compute the probabilities of {\it atypical large}
fluctuations which go beyond the range of validity of the limiting
distributions.

For $p=0$
we show that $P_0(N|E)$, as a function of $N$ for fixed $E$, has a peak at a characteristic
value $N_0^*(E)\simeq \frac{1}{a}\,\sqrt{E}\, \log(4E/a^2)$ for large $E$, where $a=\pi
\sqrt{2/3}$,
and the random variable $N$ {\it typically} fluctuates around $N_0^*(E)$ over a scale
$\sim \sqrt{E}$. Moreover, in the vicinity of $N_0^*(E)$ over a range
$|N-N_0^*(E)|\sim O(\sqrt{E})$, the distribution $P_0(N|E)$ has
a scaling form (or a limiting law). In terms of the cumulative probability,
\begin{equation}
Q_0(N|E)=\sum_{N'=0}^N P_0(N|E) \approx
F_0\left(\frac{a}{2\sqrt{E}}\,(N-N_0^*(E))\right),
\label{gumbel1}
\end{equation}
where the scaling function $F_0(z)$ has an asymmetric Gumbel form, thus recovering
the Erd\"os-Lehner result~\cite{Erdos}
\begin{equation}
F_0(z)=\exp[-\exp[-z]].
\label{gumbel2}
\end{equation}

In contrast, for $p>0$, the 
distribution $P_p(N|E)$ has
quite a different asymptotic behavior. It has a peak at a characteristic
value $N_p^*(E)\simeq
a_1(p) \sqrt{E}$ and $N$ typically fluctuates around $N_p^*(E)$ over a scale
of $\sim E^{1/4}$ for all $p$. Moreover we show that, on this scale, the fluctuations
are Gaussian. More precisely, we show that in the vicinity of $N_p^*(E)$,
the cumulative probability $Q_p(N|E)$ has a scaling form
\begin{equation}
Q_p(N|E)\approx F\left(\frac{(N-a_1(p) \sqrt{E})}{a_2(p)\,E^{1/4}}\right)\quad\quad\, 
{\rm
where}\quad\quad
F(z)=\frac{1}{\sqrt{2\pi}}\int_{-\infty}^z e^{-y^2/2}dy
\label{fermi1}
\end{equation}
is a universal scaling function independent of $p$ ($>0$). The two
nonuniversal scale factors
$a_1(p)$ and $a_2(p)$ however depend explicitly on $p$ 
and can be computed exactly.
For example, for Fermions ($p=1$), we recover the Erd\"os-Lehner result for the mean 
\begin{equation}
a_1(1)= \frac{2\sqrt{3}}{\pi}\ln (2)
\end {equation}
and  get a new result for the variance 
\begin{equation} 
a_2(1)=\left[
\frac{3\pi^2-36\,\ln^2(2)}{\sqrt{3}\, \pi^3}
\right]^{1/2}.
\label{fermi2}
\end{equation}

\begin{figure}[htbp]
\epsfxsize=8cm
\centerline{\epsfbox{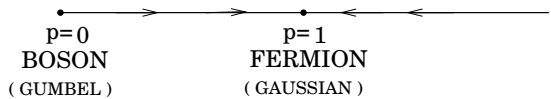}}
\caption{Schematic flows along the $p$ axis. The $p=0$ represents the
Bosonic fixed point where the limiting distribution of $P_p(N|E)$ is Gumbel. In contrast,
the behavior for all $p>0$ is controlled by the Fermionic fixed point
at $p=1$ where the limiting distribution is Gaussian.}
\label{fig:fp}
\end{figure}

Thus, as far as the limiting shape of the scaled distribution of $P_p(N|E)$ 
is concerned, it is a universal Gaussian for all $p>0$. The Fermionic case
$p=1$ is thus a representative of all $p>0$ and can be considered as an attractive
fixed point along the $p$-axis (see Fig. (\ref{fig:fp}). In contrast, the
Bosonic case $p=0$ represents a repulsive fixed point where
the shape is Gumbel.

The limit laws above describe the probabilities of {\it typical} fluctuations
of $N$ around its characteristic value $N_p^*(E)$? In this paper, we have
also investigated the probability of {\it atypical large}
fluctuations of $N$ away from $N_p^*(E)$ and calculated the
corresponding large deviation functions exactly. Like the limit laws, the
large deviation properties for $p>0$ turns out to be rather different
from the $p=0$ case, thus confirming the fixed point picture
of Fig. \ref{fig:fp}. Curiously though, we show that the large deviation function for any 
$p>0$ is related to that of $p=0$ via 
an exact nonlinear relation. 

The paper is organized as follows. In Section 2, we precisely define the MDP
problem, provide an exact derivation of the generating function of 
$\rho_p(E,N)$  
and establish a nonlinear relation between $\rho_p(E,N)$ with $p>0$ 
and $\rho_0(E,N)$. In Section 3, we show how the MDP problem with
$0<p<1$ corresponds to exclusion statistics. In Section 4, we 
provide detailed asymptotic analysis of $\rho_p(E,N)$ for all $p\ge 0$
and obtain the limiting shapes of the scaled distribution $P_p(N|E)$
and also calculate exactly the associated large deviation functions. 
Finally, we conclude with a summary and open problems in Section 5.

\section{Minimal Difference Partition Problem: A Combinatorial Approach}

\begin{figure}[htbp]
\epsfxsize=8cm
\centerline{\epsfbox{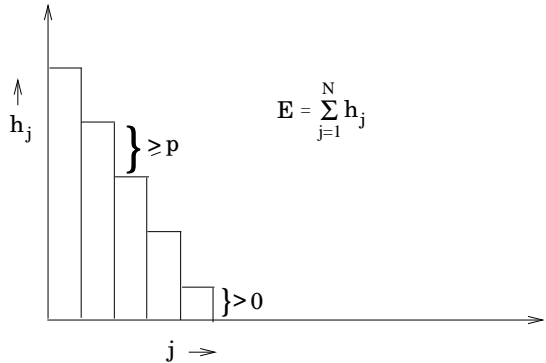}}
\caption{A typical partition configuration of the MDP problem with $N=5$.
The column heights $h_j>0$ for all $j=1,2,\ldots N$ and their total
height is $E=\sum_{j=1}^N h_j$. In addition, they satisfy the 
constraint, $(h_j-h_{j+1})\ge p$ for an integer $p$ for all
$j=1,2,\ldots ,N-1$.} 
\label{fig:mdp}
\end{figure}

In the minimal difference partition (MDP) problem, one partitions
an integer $E$ into $N$ nonzero parts, $E=\sum_{j=1}^N h_j$ (with $h_j>0$ for
each $j=1,2,\ldots N$) such that each part exceeds the next one by at least
an integer $p$, i.e., $(h_j-h_{j+1})\ge p$ for all $j=1,2,
\ldots, N-1$ (see Fig. \ref{fig:mdp}).
Let $\rho_p(E,N)$ denote the number of ways one can
achieve this. 
Clearly, the cases $p=0$ and $p=1$ reduce respectively to the unrestricted
partitions (Bosons) and the restricted partitions (Fermions). The generating
function for $\rho_p(E,N)$ is well known~\cite{Andrews2} and is given in Eq. (\ref{gen4}).
However, here we provide a simple derivation of this result that brings out
in a direct way a nontrivial connection between the cases $p>0$ and $p=0$ 
which will be used later for the analysis of the asymptotic behavior. 

Let us first establish an exact one to one correspondence between 
a partition configuration of the MDP with nonzero $p>0$ and a
partition configuration with $p=0$. Let $\{h_j\}$ denote the set
of nonzero heights in the partition of $E=\sum_{j=1}^N h_j$ for $p=0$
(Bosonic case). Thus, $h_j\ge h_{j+1}$ for all $j=1,2,\ldots, N-1$.
Let us now define a new set of heights $h_j'= h_j +p(N-j)$ for
$j=1,2,\ldots, N$. Thus, $h_j'-h_{j+1}'=h_j-h_{j+1}+p$ for all $j=1,2,\ldots, N-1$
and $h_N'=h_N>0$. The new 
heights thus satisfy the 
constraint $(h_j'-h_{j+1}')\ge p$ for all $j=1,2,\ldots, N-1$ and their
total height is given by 
\begin{equation}
E'=\sum_{j=1}^N h_j' = E+ pN(N-1)/2=\sum_{j=1}^N h_j+pN(N-1)/2.
\label{ph1}
\end{equation} 
Therefore, the primed heights correspond to a partition configuration
of the integer $E'$ into $N$ parts with $p>0$. This exact correspondence then
provides us 
with the following identity valid for all $N$
\begin{equation} 
\rho_p(E,N)= \rho_0\left(E-\frac{p}{2}\,N(N-1), N\right).
\label{iden1}
\end{equation}
Thus if one can compute the partition function $\rho_0(E,N)$ for the Bosonic case
($p=0$), this identity can be used to obtain exact results for any arbitrary $p>0$, 
including
the Fermionic case $p=1$.

For the Bosonic ($p=0$) case, a straightforward calculation gives the generating function
 \begin{equation}
 \sum_{E=1}^{\infty} \rho_0(E,N) x^E = \frac{x^N}{(1-x)(1-x^2)\ldots(1-x^N)}.
 \label{gen3}
\end{equation}
Using the correspondence in Eq. (\ref{iden1}) gives the general result for all $p\ge 0$
\begin{equation}
\sum_{E=1}^{\infty} \rho_p(E,N) x^E = \frac{x^{N+ pN(N-1)/2}}{(1-x)(1-x^2)\ldots(1-x^N)}.
\label{gen4}
\end{equation}

It turns out to be convenient sometimes to use the cumulative
partition function $C_p(E,N)=\sum_{N'=0}^N \rho_p(E,N)$. For $p=0$,
its generating function can be easily derived from Eq. (\ref{gen3}) 
and has a particularly 
simple form which turns out to be rather useful,   
\begin{equation}
\sum_{E=1}^{\infty} C_0(E,N) x^E = \frac{1}{(1-x)(1-x^2)\ldots(1-x^N)}.
\label{cumgenb}
\end{equation}
Comparing Eqs. (\ref{gen4}) and (\ref{cumgenb}) one gets another identity
\begin{equation}
\rho_p(E,N)= C_0\left(E-N-\frac{p}{2}\, N(N-1),N\right)
\label{gen5}
\end{equation}
which we will be using later.

\section{The MDP with $0<p<1$ and Exclusion Statistics}

In this section we show that the MDP problem with integer parameter $p$, continued
analytically to the range $0\le p\le 1$ corresponds to a quantum gas of
interacting particles obeying exclusion statistics. This correspondence
is established at two levels: (i) at a microscopic level where we show
in subsection 3.1 that the $\rho_p(E,N)$ of the MDP problem corresponds precisely to
the micro-canonical partition function of the one dimensional Calogero model
in an external harmonic potential 
and (ii) at
a more general thermodynamical level in section 3.2.  

\subsection{Equivalence between the MDP problem and the spectrum of the Calogero model
in a harmonic well}

The aim of this subsection is to show that there is an exact one to one correspondence
between the energy levels of the one dimensional Calogero model in an external
harmonic well and the 
partition configurations of the MDP problem, continued analytically
to $0\le p\le 1$ in the sense explained below.

The Calogero model (for a review see \cite{Pasquier}, \cite{Polychronakos}) describes an interacting quantum particle system on a line
where the particles attract each other by an inverse square potential. In order
to have a proper thermodynamic limit, one can either put the $N$ particles in
a finite box of size $L$ and then take the $N\to \infty$, $L\to \infty$ limit
keeping the density $N/L$ fixed. Alternatively, one can keep the particles
on the infinite line, but switch on an external harmonic
potential of strength $\omega$. In the latter case, one has to eventually
take the limit $\omega\to 0$ in a suitable way.
It turns out that 
while the model in a box is not integrable, the
 model in a harmonic potential is integrable. Setting the Planck's constant 
$\hbar=1$ and the mass of each particle $m=1$, the quantum 
Hamiltonian of the model
is 
\begin{equation}
{\hat H}= -\frac{1}{2}\sum_{i=}^N \frac{\partial^2}{\partial x_i^2} + 
\alpha(1+\alpha)\sum_{i<j}\frac{1}{(x_i-x_j)^2}
+\frac{1}{2}\,\omega^2\, \sum_{i=1}^N x_i^2
\label{calo1}
\end{equation}
where $x_i$ represents the position of the $i$-th particle,
$\omega$ represents the frequency of the external harmonic well and
$\alpha\in [-1,0]$ represents the coupling strength of mutually attractive
interaction between the particles. In addition, the many body wavefunction
must vanish at $x_i=x_j$ for any pair ($i\ne j$) of coordinates for $\alpha\ne 0$. 
It turns out that while $\alpha=0$ represents non-interacting Bosons (where the
many body wavefunction is symmetric under the exchange of $x_i$ and $x_j$), 
$\alpha=-1$ represents noninteracting Fermions (where the wavefunction vanishes
at $x_i=x_j$). For other values of $\alpha$, this model is known
to exhibit 
fractional statistics (see in particular in the next subsection its manifestation in the thermodynamics of the model).

The many-body energy spectrum
of this model is known exactly~\cite{Calogero}. The energy $E(\{h_j\})$ is labelled
by non-increasing integers $h_1\ge h_2\ge h_3\ldots \ge 1$
\begin{equation}
E(\{h_j\})= \omega\left[\sum_{j=1}^N h_j -\frac{1}{2}\,\alpha\, N(N-1)\right].
\label{spec1}
\end{equation}
By making a shift as in Eq. (\ref{ph1}), i.e., defining a new set of
variables $h_j'=h_j+ \alpha(N-j)$, 
one can express the energy as 
$E= \omega \sum_{j=1}^N h_j'$ with the constraint that
$(h_j'-h_{j+1}')\ge -\alpha$. Thus the spectrum of the Calogero model in
a harmonic potential corresponds
exactly to partition configurations of the MDP with parameter $p=-\alpha$, but now $p$ is a 
real number such that
$0\le p\le 1$. Hence, the micro-canonical partition function
of the Calogero model $\rho_{\rm Cal}(E,N)$ denoting the number of 
configurations
with energy $E$ and number of particles $N$ is directly related to
the number of partitions $\rho_p(E,N)$ of the MDP model via
\begin{equation}
\rho_{\rm Cal}(E,N) = \rho_p(E/\omega, N)
\label{micro1}
\end{equation}
This implies that the grand-canonical partition functions of the two models
are also related. Let $Z_{\rm Cal}(\beta, z)= \sum_{E,N} \rho_{\rm Cal} (E,N) e^{-\beta\, E} 
\,z^N$
be the grand-canonical partition function in the Calogero model in a harmonic well of 
frequency $\omega$,
where $\beta$ is the inverse 
temperature and $z$ is the fugacity. Similarly, we define $Z_p(\beta, z)=
\sum_{E,N} \rho_p(E,N) e^{-\beta\, E}\, z^N$ as the double generating function in the MDP 
problem with parameter $p$.
The relation in Eq. (\ref{micro1}) then translates into the following
relation between the grand partition functions
\begin{equation}
Z_{\rm Cal}(\beta, z) = Z_p(\omega \beta, z).
\label{grand1}
\end{equation}

\subsection{Thermodynamic equivalence to exclusion statistics} 

Exclusion statistics can be most conveniently defined in the following
thermodynamical sense. Let $Z(\beta, z)$ denote the grand partition
function of a quantum gas of particles at inverse temperature $\beta$
and fugacity $z$. Such a  gas is said to obey exclusion statistics with parameter $0\le p\le 1$ if
$Z(\beta, z)$ can be expressed as an integral representation
\begin{equation}
\ln Z(\beta,z)=\int_0^{\infty} \rho(\epsilon)\ln y_p(ze^{-\beta\epsilon})d\epsilon
\label{thermo}
\end{equation}
where $\rho(\epsilon)$ denotes an effective single particle density
of states and the function $y_p(x)$, which encodes fractional statistics, 
is given by the solution of the functional equation
\begin{equation}
y_p(x)- x\, y_p^{1-p}(x)=1.
\label{func1}
\end{equation}
Note that for $p=0$, one gets $y_p(x)=1/(1-x)$ and for $p=1$, $y_p(x)=(1+x)$.
In these two extreme cases, Eq. (\ref{thermo}) reduces to the standard
grand partition functions of noninteracting Bosons and Fermions respectively.
The fractional statistics with parameter $0< p< 1$ (that corresponds to
an interacting gas) then smoothly 
interpolates
between these two extreme cases.

There are at least two microscopic quantum models whose grand-canonical
functions have the form of  Eq. (\ref{thermo}).
The first example is the LLL anyon model~\cite{LLLanyons} in the infinite volume limit 
which can be shown to satisfy Eq. (\ref{thermo}) with an
effective density 
of states $\rho(\epsilon)= \frac{B\, V}{\phi_0}\, 
\delta(\epsilon-\omega_c)$ where $B$ is the external magnetic field, $\phi_0=2\pi/e$
is the flux quantum, $\omega_c= eB/2m$ is the cyclotron frequency and $V$
is the infinite area of the system. In this model, the parameter $p=\phi/\phi_0$ corresponds to
the the flux carried by each anyon in units of the flux quantum. The
second example corresponds to the one dimensional Calogero model
defined in Eq. (\ref{calo1}) again in the infinite box limit.
In this case, one can show that the grand partition function again can be 
written 
in the form as in Eq. (\ref{thermo}) with an effective single
particle density $\rho(\epsilon)= L/\sqrt{8 \pi^2 \epsilon}$ 
where $L$ is the infinite length of the system.       
In both cases, the thermodynamics is computed in the presence of a 
long distance harmonic well regulator,  and the thermodynamic limit 
where the external
frequency $\omega \to 0$ is taken in such a  way so that one correctly 
recovers the infinite box limit.

Here we show, using the equivalence to the MDP problem in Eq. (\ref{grand1}),
that the grand partition
function of the one dimensional Calogero model in an external harmonic
well of frequency $\omega$, in the
limit $\omega\to 0$ can again be written
in the general form as in Eq. (\ref{thermo}), but now with an effective constant density of
states $\rho(\epsilon)= 1/\omega$.
Note that this is different from the Calogero 
model in a infinite box of size $L$ (the second example mentioned in the
preceding paragraph):
here, the particles
are sitting inside
a harmonic well with almost vanishing but non zero frequency. 

To proceed, we first calculate the grand partition function of the
MDP problem, $Z_p(\omega \beta, z)= \sum_{E,N} \rho_p(E,N) e^{-\omega \beta E} z^N$, 
starting from Eq. (\ref{gen4}). We set $x=e^{-\beta \omega}$ in Eq. (\ref{gen4}), multiply 
it by $z^N$ and sum over $N$. Next we take the logarithm on both sides
and then make a cluster expansion, $\ln Z_p(\omega \beta, z)= \sum_{n=1}^{\infty} b_n z^n$.
Now, taking the $\omega \to 0$ limit (keeping $\beta$ fixed), one gets 
\begin{equation} 
b_1\simeq {1\over \omega \beta}e^{-\omega \beta} \quad \quad 
b_{n\ge 2}\simeq {1\over 
\omega \beta}{e^{-n\omega \beta}\over n^2} \prod_{k=1}^{n-1}(1-{p n\over k})
\label{cluster1}
\end{equation}
Consequently, since from (\ref{func1}) 
$\ln y_p(x)=
\sum_{n=1}^{\infty}\frac{x^n}{n}\, 
\prod_{k=1}^{n-1}(1-{p n\over k})$,
one infers that, provided the series is convergent, that is 
$ze^{-\omega \beta}<1$ 
\begin{equation}
\ln Z_{p}(\omega \beta,z)=\int_1^{\infty} \ln y_p(ze^{-\omega \beta\epsilon})d\epsilon.
\label{cluster2}
\end{equation}
Making a further change of variable $\omega \epsilon\to \epsilon$, it follows that
in the limit $\omega \to 0$
\begin{equation}
\ln Z_{p}(\omega \beta,z) \to \frac{1}{\omega} \int_0^{\infty}  \ln y_p(ze^{- 
\beta\epsilon})d\epsilon.
\label{cluster3}
\end{equation} 
This is again of the form in Eq. (\ref{thermo}) with $\rho(\epsilon)=1/\omega$.
Using the equivalence in Eq. (\ref{grand1}) we then conclude
that the Calogero model in an external harmonic well with
vanishing frequency, which precisely corresponds to the MDP problem with
parameter $0\le p\le 1$, can be viewed 
as a gas of particles obeying exclusion statistics with a
statistical parameter $ \alpha=-p$ and a constant density of states.
 
\section{Partition Asymptotics in MDP with $p\ge 0$}

In this section we compute explicitly the asymptotics of the probability 
distribution $P_p(N|E)$ in the MDP problem for all $p\ge 0$. We show that while 
the
limiting shape of this distribution (properly centered and scaled)
is Gumbel for $p=0$, it is Gaussian for all $p>0$ including the
Fermi case $p=1$.

\subsection{ Bosonic case $p=0$}

Our starting point is the generating function for the $C_0(E,N)$ in Eq. (\ref{cumgenb}).
We formally invert this generating function using Cauchy's theorem and write
\begin{eqnarray}
C_0(E,N)&= &\frac{1}{2\pi i}\int \frac{dx}{x^{E+1}} 
\frac{1}{(1-x)(1-x^2)\ldots(1-x^N)}\nonumber \\
&=& \frac{1}{2\pi i}\int d\beta\, \exp\left[\beta E -\sum_{k=1}^N \ln\left(1-e^{-\beta 
k}\right)\right],
\label{cont1}
\end{eqnarray}
where the integration is in the complex $x$ plane along a contour around the origin and
we have made a change of variable $x=\exp(-\beta)$ in going to the second line.
For large $E$, one can then analyze the leading asymptotic behavior by employing
the saddle point method in the complex $\beta $ plane. Anticipating that for large $E$,
the most important contribution to the integral will come from small $\beta$, we first
obtain the leading small $\beta$ behavior of the action $S_{E,N}(\beta)= \beta E
-\sum_{k=1}^N \ln\left(1-e^{-\beta k}\right)$. Using the Euler-Mclaurin summation formula,
one can easily show that in the limit of $\beta\to 0$, $N\to \infty$ limit but keeping
$\beta N$ fixed, the action can be written as
\begin{equation}
S_{E,N}(\beta) \simeq \beta E + \frac{1}{\beta}\int_0^{\beta N} \frac{t\, dt}{e^t -1} -N 
\ln\left(1-e^{-\beta N}\right).
\label{action1}
\end{equation}
We next maximize the action with respect to $\beta$, i.e., we set ${\partial S}/{\partial 
\beta}=0$ to get
\begin{equation}
E= \frac{1}{\beta^2} \int_0^{\beta N} \frac{t\, dt}{e^t -1}.
\label{saddle1}
\end{equation}
For a given large $E$, one gets $\beta^*$ by implicitly solving the saddle point equation
(\ref{saddle1}) and substitute it back in the action $S_{E,N}(\beta^*)$. Thus, to
leading order,
\begin{equation}
C_0(E,N) \simeq \exp\left[S_{E,N}(\beta^*)\right]
\label{saddle2}
\end{equation}
where the micro-canonical entropy $S_{E,N}(\beta^*)$ can be written as 
\begin{equation}
S_{E,N}(\beta^*)= \frac{1}{\beta^*}\left[2 \int_0^{\beta^* N} \frac{t\, dt}{e^t -1} -\beta^*N
\ln\left(1-e^{-\beta^* N}\right)\right].
\label{entropy1}
\end{equation}

To bring out the scaling form of $C_0(E,N)$ explicitly for large $E$ and $N$, we
next proceed as follows.
It is evident from the structure of the saddle point solution that $\beta^*\sim 
E^{-1/2}$
for large $E$, whereas $\beta^*\sim 1/N$ for large $N$ 
indicating that the correct scaling variable
is $x=N/\sqrt{E}$. Next we set $\beta^* N= H(x)$. In terms of these new scaling variables,
the saddle point solution in Eq. (\ref{saddle1}) can be recast as
\begin{equation}
\frac{H^2(x)}{x^2}= \int_0^{H(x)}\frac{t\, dt}{e^t-1}.
\label{scaling1}
\end{equation}
Thus, given $x$, one has to find $H(x)$ by implicitly solving Eq. (\ref{scaling1}).
The entropy in Eq. (\ref{entropy1}) becomes
$S_{E,N}(\beta^*)= \sqrt{E}\, g(x)$ where the scaling function $g(x)$
is given from Eq. (\ref{entropy1}) as
\begin{equation}
g(x) = 2 \frac{H(x)}{x}- x \ln\left(1-e^{-H(x)}\right).
\label{scaling2}
\end{equation}
Thus, asymptotically for large $N$ and $E$, keeping the ratio $x=N/\sqrt{E}$ fixed,
the cumulative number of configurations $C_0(E,N)$ for Bosons can be
written as
\begin{equation}
C_0(E,N)\simeq \exp\left[\sqrt{E}\, g\left(\frac{N}{\sqrt{E}}\right)\right]   
\label{ldvcb}
\end{equation}
where $g(x)$ is the large deviation function given exactly by Eqs. (\ref{scaling2})
and (\ref{scaling1}). This is the main result of this subsection. 

\begin{figure}[htbp]
\epsfxsize=8cm
\centerline{\epsfbox{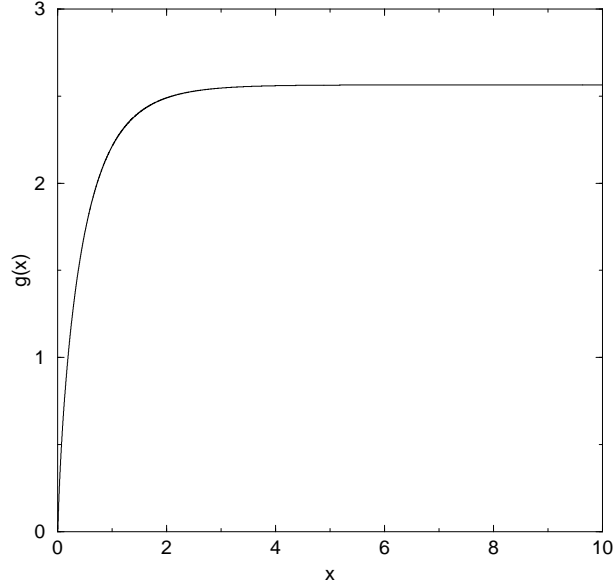}}
\caption{The large deviation function $g(x)$ for Bosons ($p=0$).}
\label{fig:gx}
\end{figure}

The function $g(x)$ has to be determined numerically by solving the implicit equations 
(\ref{scaling2})
and (\ref{scaling1}). A plot of this function is given in Fig. (\ref{fig:gx}). The
asymptotic properties of $g(x)$ for small and large $x$ can be worked out easily.
It can be shown that
\begin{eqnarray}
g(x)&\approx & -2x \ln (x) \quad\quad {\rm as}\quad x\to 0 \nonumber \\
&\approx & a - \frac{2}{a}\,\exp(-a x/2) \quad\quad {\rm as}\quad x\to \infty
\label{asgx}
\end{eqnarray}
where $a= \pi \sqrt{2/3}= 2.5651\dots$. 

The result $g(x\to\infty)=a$ implies, from Eq. (\ref{ldvcb}) that $\rho(E)=C(E, N\to 
\infty)\sim \exp[a \sqrt{E}]$ to leading order for large $E$, thus recovering
the famous Hardy-Ramanujan result\cite{HR}. The normalized cumulative 
distribution
of $N$ (given $E$), $Q_0(N|E)=C_0(E,N)/\rho(E)$ then has the large deviation 
form
\begin{equation}
Q_0(N|E)\simeq \exp\left[-\sqrt{E} \Phi\left(\frac{N}{\sqrt{E}}\right)\right]\quad\quad
{\rm where}\quad \Phi(x)= a - g(x)
\label{phix1}
\end{equation}
and $\Phi(x)$ has the asymptotic behavior
\begin{eqnarray}
\Phi(x) &\approx & a + 2x \ln (x) \quad\quad {\rm as}\quad x\to 0 \nonumber \\
&\approx & \frac{2}{a}\,\exp(-a x/2) \quad\quad {\rm as}\quad x\to \infty. 
\label{phix2}
\end{eqnarray}

As $x\to \infty$, i.e., as $N>> \sqrt{E}$, clearly
$Q(N|E)\to 1$ as expected, since it is the normalized cumulative
distribution of $N$. The precise approach to $1$ can be obtained
using the large $x$ asymptotics of $\Phi(x)$ in Eq. (\ref{phix2}).
Substituting this behavior in Eq. (\ref{phix1}) one gets for
$N>> \sqrt{E}$,
\begin{equation}
Q_0(N|E)\simeq \exp\left[-\frac{2}{a}\sqrt{E}\, \exp(-aN/{2\sqrt{E}})\right]
=F_0\left(\frac{a}{2\sqrt{E}}\,(N-N_0^*(E))\right),
\label{gumb1}
\end{equation}
where the characteristic value of the random variable $N$ is $N_0^*(E)\simeq 
 \frac{1}{a}\,\sqrt{E}\, \log(4E/a^2)$ and the scaling function has the
Gumbel form, $F_0(z)= \exp[-\exp[-z]]$. Evidently, the probability
distribution $P_0(N|E)=Q(N|E)-Q(N-1|E)\simeq {\partial Q_0(N|E)}/{\partial N}$
has the scaling form
\begin{equation}
P_0(N|E)\simeq 
\frac{a}{2\sqrt{E}}F_0'\left(\frac{a}{2\sqrt{E}}\,(N-N_0^*(E))\right)\quad\quad
{\rm where}\quad  F_0'(z)= \exp[-z-\exp[-z]]
\label{pdfgumbel}
\end{equation}
which is highly asymmetric around the peak at $N=N_0^*(E)$. This limiting
distribution of $N$ that describes the probability of {\it typical}
fluctuations of $N$ of $\sim O(\sqrt{E})$ around the peak at 
$N_0^*(E)$,  
was originally derived by Erd\"os and Lehner by computing upper
and lower bounds to the probability\cite{Erdos}. Our method allows
us to obtain a more general result in Eq. (\ref{phix1}) which is valid
over a wider range and reduces to the Gumbel limiting form near the peak. A rigorous mathematical 
derivation of this result, including the exponential prefactor, can be found in the work of Szekeres \cite{Szekeres}.

\subsection{The case $p>0$}

For $p>0$, one can directly obtain the asymptotic behavior of
$\rho_p(E,N)$ by using the identity in Eq. (\ref{gen5}) and
the already derived asymptotic behavior of $C_0(E,N)$ in Eq. (\ref{ldvcb}).
In the scaling limit when $N$ and $E$ are both large but the ratio $x=N/\sqrt{E}$
is kept fixed, one gets to leading order
\begin{equation}
\rho_p(E,N)\simeq \exp\left[\sqrt{E}\, f_p\left(\frac{N}{\sqrt{E}}\right)\right]
\label{p1}
\end{equation}
with
\begin{equation}
f_p(x) = \sqrt{1-px^2/2}\,\, g\left(\frac{x}{\sqrt{1-px^2/2}}\right)
\label{fpx}
\end{equation}
where $g(x)$ is given in Eqs. (\ref{scaling2}) and (\ref{scaling1}). 

\begin{figure}[htbp]
\epsfxsize=8cm
\centerline{\epsfbox{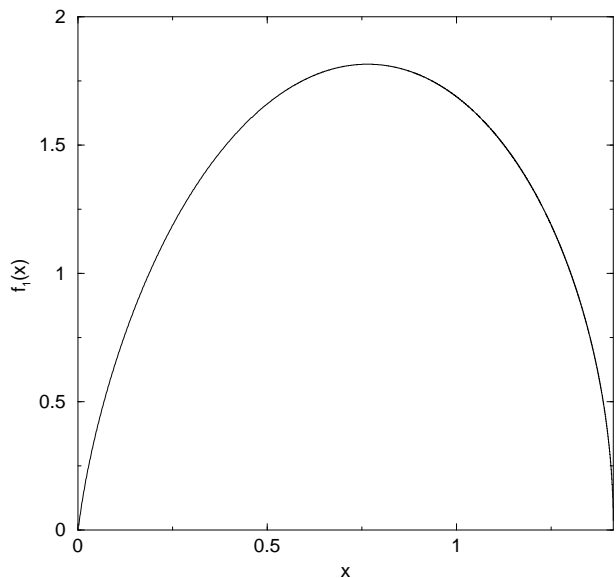}}
\caption{The large deviation function $f_1(x) $ for Fermions ($p=1$).}
\label{fig:f1x}
\end{figure}

Note 
that the function $f_p(x)$ has nonzero support only over $x\in \left[0, 
\sqrt{2/p}\right]$.
This is easy to understand from the fact that for $p>0$, $E$ has a minimum
value for any given $N$, or equivalently $N$ has a finite 
maximum 
value for any given $E$. For example, for the Fermionic case ($p=1$),  
the lowest value of $E$ for a given $N$ corresponds to the Fermi energy
$E_{F}= N(N+1)/2$ where one puts one Fermion at each single particle level
$\epsilon_i=i$ for $i=1,2,\ldots ,N$. Thus, $E\ge N(N+1)/2$ for all $N$.
In other words, for large $N$, $N\le \sqrt{2E}$, i.e., $x\le \sqrt{2}$.
Similar arguments can be given for any positive $p>0$. Unlike the function $g(x)$
which is monotonically increasing, the function $f_p(x)$ in Eq. (\ref{fpx})
is a non-monotonic function in $x\in \left[0,\sqrt{2/p}\right]$. It vanishes at
the two ends as
\begin{eqnarray}
f_p(x) & \approx & -2x\, \ln(x)\quad\quad {\rm as}\quad x\to 0 \nonumber \\
&\approx & \frac{\pi \sqrt 6}{3} (2p)^{1/4}\sqrt
{\sqrt {2/p} -x}
\label{asymfp}
\end{eqnarray} 
and has a unique maximum at $x^*(p)=a_1(p)$, where $a_1(p)$
can be obtained by setting $df_p(x)/dx=0$ in Eq. (\ref{fpx}) and then using
the known properties of $g(x)$. A plot of $f_p(x)$ for $p=1$ (Fermi case)
is given in Fig. (\ref{fig:f1x}).

By playing around with the form of $f_p(x)$ in Eq. (\ref{fpx}) and that of
$g(x)$ in Eqs. (\ref{scaling2}) and (\ref{scaling1}) one can derive a
number of explicit results. We skip the details here and just mention
the results. For example, the location of the maximum $x^*(p)=a_1(p)$
is given by
\begin{equation}
a_1(p)= \frac{\ln y_0^\star}{\sqrt{p(\ln y_0^\star)^2/2+Li_2(1-1/y_0^\star)}}
\label{a1p}
\end{equation}
where $ y_0^\star -
y_0^{\star {1-p}}=1$ and $Li_2(z)= \sum _{k=0}^\infty \frac {z^k}{k^2} $ is the dilogarithm 
function. For example, in the Fermionic case $p=1$, we get $y_0^\star=2$ and
$a_1(1)= 2\sqrt{3} \ln(2)/\pi=0.764304\ldots $. Similarly, the value of the function at the 
maximum
$f_p(x=a_1(p))$ can be shown to be
\begin{equation}
f_p(x=a_1(p)) =  2 \sqrt {Li_2 (1-1/y_0^\star)+p (\ln y_0^\star)^2/2 }
\label{fpmax}
\end{equation}
For example, for $p=1$, it gives $f_p(x=a_1(1))= \pi/\sqrt{3}$. 
For an arbitrary $p$, this formula which goes back to Meinardus \cite{Meinardus} , provides a generalization of the Hardy-Ramanujan formula for $\rho(E)$. One can check that  Eq.(\ref{fpmax} ) co\"incides with the integral formula obtained by Blencowe et al \cite{Blencowe} who also investigated the Haldane statistics but didn't provide any combinatorial interpretation of their result.
Around the maximum at $x=a_1(p)$, the function $f_p(x)$ can be expanded in 
a Taylor series and up to the quadratic order 
\begin{equation}
f_p(x) \simeq f_p(a_1(p)) + \frac{1}{2 a_2^2(p)}\, (x- a_1(p))^2 +\ldots
\label{quad1}
\end{equation}
where $a_2(p)$ can also be evaluated. For example, for $p=1$, we get
\begin{equation}
a_2(1)= \left[\frac{3\pi^2-36\,\ln^2(2)}{\sqrt{3}\, \pi^3}\right]^{1/2}=0.478815\ldots
\label{a21}
\end{equation}

Evidently, one can easily evaluate the asymptotic 
behavior of $\rho_p(E)=\sum_{N} \rho_p(E,N)$ for large $E$ by replacing
the sum by an integral, use
the large deviation form in Eq. (\ref{p1}) for $\rho_p(E,N)$ and then
using the saddle point method. To leading order, this gives 
\begin{equation}
\rho_p(E) \simeq \exp\left[f_p(a_1(p))\, \sqrt{E}\right].
\label{rhope}
\end{equation}
The normalized probability distribution of $N$ (for fixed $E$) $P_p(N|E)= 
\rho_p(E,N)/\rho_p(E)$
then has the large deviation asymptotics
\begin{equation}
P_p(N|E)\simeq \exp\left[- \sqrt{E}\,\psi_p\left(\frac{N}{\sqrt E}\right)\right]\quad \quad
{\rm where} \quad \psi_p(x) = f_p(a_1(p))-f_p(x)
\label{psipx}
\end{equation}

Thus for all $p>0$, the probability distribution $P_p(N|E)$ has a peak at 
a characteristic value $N_p^*(E)= a_1(p) 
\sqrt{E}$ (note the  
difference from the Boson case $p=0$ where $N_0^*(E)\sim \sqrt{E} \ln (E)$).
Using the expansion in Eq. (\ref{quad1}) it follows that
in the vicinity of $N_p^*(E)$ (over a scale of $\sim O(E^{1/4})$), $P_p(N|E)$ has a 
Gaussian limiting form
\begin{equation}
P_p(N|E)\simeq \frac{1}{a_2(p) E^{1/4}} F'\left(\frac{(N-a_1(p) 
\sqrt{E})}{a_2(p)\,E^{1/4}}\right)\quad\quad\,
{\rm
where}\quad\quad
F'(z)=\frac{1}{\sqrt{2\pi}} e^{-z^2/2}
\label{fermigauss}
\end{equation}
Note in particular, that the standard deviation measuring the root mean square fluctuation
of $N$, $\sigma_p(E)= \sqrt{\langle (N- N_p^*(E))^2\rangle}$ grows with $E$ as a
power law, $\sigma_p(E)\simeq a_2(p) E^{1/4}$ where the exponent $1/4$ is
universal for all $p>0$. 
Moreover, apart from nonuniversal $p$ dependent
scale factors such as $a_1(p)$ and $a_2(p)$, 
the full distribution $P_p(N|E)$ also has the same universal
Gaussian limiting form for all $p>0$. Thus, the Fermi point $p=1$
is a generic point that is representative of all values of $p>0$ as
far as the limiting distribution is concerned. In this sense, all $p>0$ behavior
is controlled by the attractive Fermi fixed point as shown in Fig. (\ref{fig:fp}).
The Bosonic fixed point at $p=0$, on the other hand, is a repulsive one. 

\section{Summary and open problems} 

To summarize, in this paper we have provided a combinatorial 
interpretation of exclusion statistics in terms of minimal difference partitions(MDP). 
This correspondance is based on the observation that the grand-canonical partition 
function of 
the Calogero model coincides with the generating function of MDP. By going to the 
grand-canonical ensemble and taking a suitable thermodynamic limit, we have recovered 
the functional equation characteristic of exclusion statistics. Apart from establishing 
this correspondance, we have also provided a detailed analysis of the asymptotic 
behaviour of $\rho_p(E,N)$. Our approach uses a mapping with the bosonic problem which 
holds for arbitrary $p\in [0,1]$. In physical terms this generalises the well known 
mapping between fermions and bosons with a linear dispersion 
law \cite{Anghel,Patton}. The fact that this mapping has a number theoretical 
interpretation was apparently not known before. By using this mapping, we obtain a 
general description of the limiting laws of $P_p(N|E)$ for all $p >0$. We find that the 
bosonic point is a repulsive fixed point where the statistics is Gumbel. In contrast for 
all $ p>0$, the distribution is Gaussian. Several questions emerge from this work and 
would be worth investigating.

1) The regime $p<0$. In this case the functional equation (\ref{func1}) still holds 
where $y_p (x)$ can be interpreted 
as the generating function of connected clusters on a p-ary tree. 
A preliminary investigation of this model shows that 
the scaling behaviour of $\rho_p(E,N)$ is quite different from the previous case~\cite{CMO}.

2) In this work we have limited ourselves to the integer partition problem or equivalently to
a quantum gas of particles with equidistant single particle spectrum, i.e., with a constant
density of states $\rho(\epsilon)={\rm const.}$.  
It would be interesting to investigate general partitions of the 
form $E=\sum n_{i}i^{s}$ that
corresponds to having a power-law density of states, 
$\rho(\epsilon)\sim \epsilon^{1/s-1}$. In this case we 
have shown in a recent work~\cite{CLM} that the bosonic sector 
gives rise to the three universal distribution laws of 
extreme statistics, namely the Gumbel, Weibull and Fr\'echet distributions. 
It would be interesting to explore the 
general $p>0$ case including the fermionic sector  
and see if the bosonic point $p=0$ is still a repulsive fixed point.

3) For the bosonic case $(p=0$), 
Vershik~\cite{Vershik} and Temperley \cite{Temperley} calculated the limiting shapes of the 
Young diagram, i.e., the average height profile for a fixed but large $E$. This result can 
be
generalized~\cite{CMO} to the case $p>0$ using the functional equation
(\ref{func1}).

\end{document}